**Title:** STATIC STRUCTURAL BEHAVIOUR OF WIRE BEARINGS UNDER AXIAL LOAD: COMPARISON WITH CONVENTIONAL BEARINGS AND STUDY OF DESIGN AND OPERATIONAL PARAMETERS

**Authors:** Iñigo Martín[1], Iker Heras[1], Josu Aguirrebeitia[1], Mikel Abasolo[1], Ibai Coria[1]

1   Department of Mechanical Engineering, Faculty of Engineering Bilbao, University of the Basque Country (UPV/EHU), Plaza Ingeniero Torres Quevedo, 1, 48013, Bilbao, Spain.

[*] Corresponding author:

  Iñigo Martín

  E-mail address: inigo.martin@ehu.eus

  Tel.: +0034 606 061 117

## Abstract

In wire bearings the rolling process occurs on raceways machined on steel wires, and the rings are made of light materials such as aluminium. This particular architecture provides both weight and inertia savings, but also significantly different behaviour with respect to conventional bearings. For this reason, specific design and analysis tools must be developed; as a first step, this work uses Finite Element models to study the influence of different parameters on the static structural response of wire bearings. Thus, bearing stiffness, load capacity and contact status (contact force and angle, and ellipse truncation) have been evaluated for several combinations of conformity, friction coefficient and boundary conditions. The results have been compared with an equivalent conventional bearing, shedding light on the main structural features of wire bearings.

## Keywords

Wire Bearing; Slewing Bearing; Stiffness; Conformity; Capacity; Truncation.

## 1. Introduction

Slewing bearings are used in slow-turn heavy-duty working conditions due to their capacity to face external axial and radial loads as well as tilting moments. Thus, they are used to slowly rotate structural elements and transfer the loads to the main structure. In conventional slewing bearings, widely used in areas such as construction machinery, renewable energies or machine tool, rolling elements are in direct contact with the bearing rings, which are made of steel. Fig. 1(a) shows the cross-section of a four-point contact slewing bearing. Extensive research has been published about the structural behaviour of these bearings in terms of static load carrying capacity [1-8], stiffness [9,10] and friction torque [11-14], among other research topics.

In 1936, Erich Franke developed and patented a new concept of bearing, the wire bearing [15]. Wire bearings are a further development of conventional slewing bearings, where the raceway is shaped in a wire located between the rolling element and the ring, as illustrated in Fig. 1(b) for the case of four-point contact bearings. This modification allows building the rings and the wires with different materials; the wires can be manufactured with hardened steel and the rings with lighter materials (aluminium, composites, plastics…). The choice of a lighter material for the rings involves significant weight savings (up to 65% according to [15]) and consequently an inertia reduction. Apart from that, wire bearings with aluminium rings have a good performance absorbing shock loads and travelling vibrations due to the lower elasticity modulus, which leads to a reduction of brinelling and striation in the raceways [16]. For these reasons, wire bearings are used where weight and inertia savings are a key aspect, such as in medical, aeronautical or military applications, among others. Little research has been published about wire bearings: Shan *et al.* [17] developed an analytical model for determining the preload in wire bearings with an unusual non-conformal contact design; Gunia and Smolnicky [18] studied the influence of certain geometrical parameters in the stress distribution along the contact with conformal wires, which is a more realistic representation of the bearings used in the industry.

This work tries to shed light on the performance of wire bearings through a comparison with well-known four-point contact slewing bearings. For that purpose, two parametric Finite Element models were created and equally loaded. As a result, direct comparisons of the static axial load capacity, bearing axial stiffness and contact behaviour were made, thus providing a global view of the advantages and shortcomings of this kind of bearings. The effect of different design and operational parameters such as the contact conformity, the friction coefficient and the stiffness of the supporting structures were considered in the study.

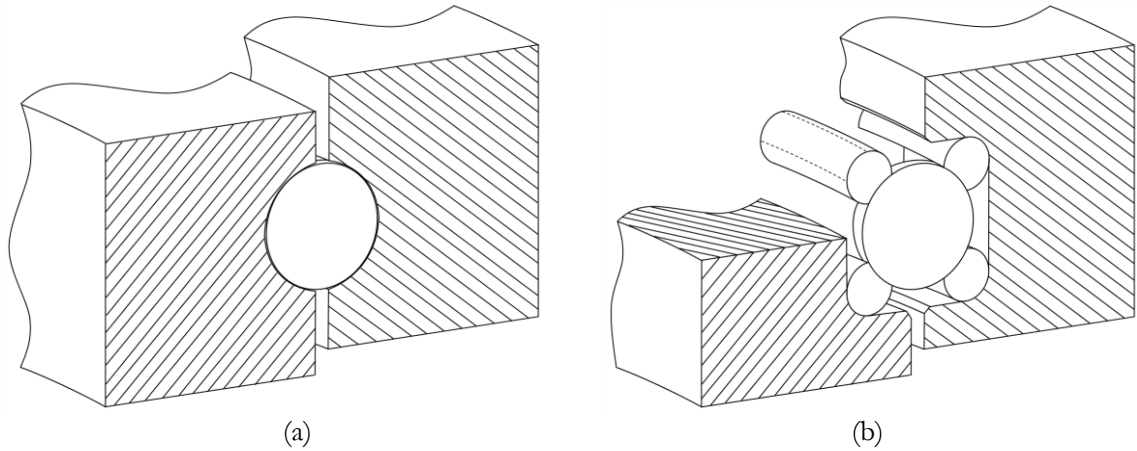

**Fig. 1.** Cross-section of a four-point contact slewing bearing: (a) conventional bearing (b) wire bearing (dashed lines are the contact lines).

## 2. Methodology

Among all the possible load cases, pure axial load case is studied. Under axial load, all the relevant phenomena susceptible of analysis appear, as wire twisting, contact ellipse truncation, effect of lubricant presence, two-point contact condition, amongst others that will be later explained. Ball preload is not considered in this work because complex interrelations were foreseen in combination with axial load. In this sense, the simplicity of the load case enables a reasonable comparison between the performances of the two bearing types. For this purpose, two parametric Finite Element (FE) models were created and several cases were raised in order to study the influence of different factors and assumptions in the performance of the bearings.

The main matter under study in the wire bearing is the ball-wire contact, because the extent of the raceway is smaller than the one in conventional slewing bearings, which reduces the surface where the hertzian elliptical contact surface can be placed. The wire-ring contact is not so critical because, as illustrated in Fig. 1(b), the contact forces are distributed along two contact lines, generating a lower contact pressure. Nevertheless, this work will prove that the behaviour of the wire-ring contact has a great influence on the performance of the wire bearing.

### 2.1. Description of the case studies

Since the aim of the work was to study and compare the performance of wire bearings with conventional bearings, different analyses were carried out varying two characteristic parameters: the osculation ratio ($s$) and the friction coefficient ($\mu$). The osculation ratio ($s$) is the main geometrical factor that defines the contact between ball and raceway; typical values close to 0.943 are used for conventional bearings [15] and between 0.87 and 0.96 for wire bearings in the industry, i.e. wire bearings tend to have a less conformal contact than conventional bearings. Regarding the friction coefficient, 0.1 is a typical value for the ball-raceway lubricated steel-steel contact pair [12,20,21]; for wire bearings, 0.1 was also used for ball-wire contact, and for wire-ring aluminium-steel contact two values were studied, 0.1 and 0.3, to evaluate the effect of the presence or absence of lubrication in the performance of the wire bearing. The first columns of Table 1 summarize the five cases analysed in this work, with their corresponding parameter values.

Even for the same type of bearing (conventional or wire), the axial stiffness is not the same for different values of $s$ and $\mu$. For this reason, to make the comparison feasible, a different displacement providing the same axial reaction force had to be applied to each model. This axial force was chosen to be the axial static load capacity as calculated by the analytical model proposed by Aguirrebeitia *et al.* [7], summarized in the last column of Table 1. The analytical model, deeply explained and validated in [7], is based on the calculation of the ball-raceway interference field caused by axial, radial and tilting displacements of the rings due to external loads (in addition to ball preload), assuming rigid rings. As the stiffness of the adjacent structures has a relevant influence in the behaviour of the bearings, two extreme situations were taken into account for each case study in Table 1: on the one hand, clamped rings, assuming that the rings are fixed to rigid supporting structures; on the other hand, unclamped rings, assuming that the supporting structures are rigid but the bearing rings can freely deform in the radial deformation). Of course, real systems behaviour is placed between these two extreme conditions.

**Table 1**
Cases under study for clamped and unclamped conditions.

| Case | Bearing type | s | μ (ball-wire) | μ (wire-ring) | $C_{0a}$ (Target axial force [kN]) [7] |
|---|---|---|---|---|---|
| 1 | **Conventional** | 0.943 | 0.1 | - | 1213.1 |
| 2 | **Conventional** | 0.870 | 0.1 | - | 674.24 |
| 3 | **Wire** | 0.870 | 0.1 | 0.1 | 674.24 |
| 4 | **Wire** | 0.943 | 0.1 | 0.1 | 1213.1 |
| 5 | **Wire** | 0.870 | 0.1 | 0.3 | 674.24 |

Regarding the geometry of the bearings, the two main geometrical parameters are the ball diameter ($D_w$) and the bearing mean diameter ($D_{pw}$), which were chosen in such a way that the resulting bearing could be found in both conventional and wire bearing commercial catalogues. The geometry of the cross-section was then obtained from [21], which proposes a standard parametric geometry for conventional four-point contact slewing bearings in terms of ($D_w$) and ($D_{pw}$). Even though wire bearing cross-sections are larger than conventional bearing sections for given ($D_w$) and ($D_{pw}$) values, the same cross-section was adopted for both bearing types in order to make the comparison in strictly the same conditions. Fig. 2 shows the cross-section of both bearings with their dimensions. The total number of balls in the bearing is n=82, and initial contact angle is α=45°.

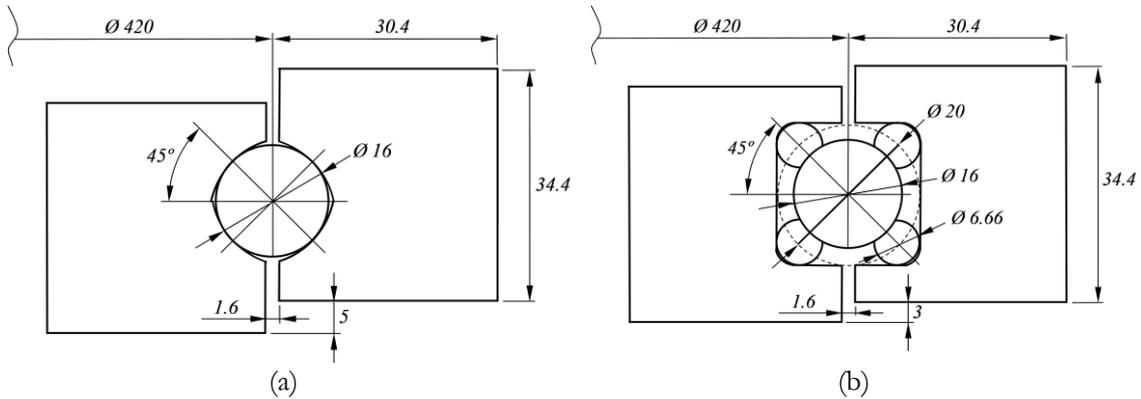

**Fig. 2.** Cross-sections of the studied bearings in [mm]: (a) conventional bearing (b) wire bearing.

## 2.2. FE Models

Two FE models were created in Ansys® to simulate the performance of the bearings, one for the conventional bearing and the other one for the wire bearing. Bolt holes were not modelled. According to the current commercial bearings, steel was used for conventional bearings (linear elastic, E=200 GPa), and for wire bearings steel was chosen for balls and wire and aluminium (linear elastic, E=71GPa) for the rings. As it will be explained next, two types of FE models were developed: half sector models and submodels.

### 2.2.1. Half sector models

The axial load situation provides a cyclic symmetry load distribution which, together with a cyclic symmetry geometry, allows to simplify the whole bearing model into a one sector model. Furthermore, the one sector model has a symmetry plane that allows analysing only one half, significantly decreasing the number of Degrees of Freedom (DoF) of the model.

Next, the models were meshed with the same element size in order to make more accurate comparison between model results. For this purpose, several partitions were carried out in the geometry. The partitions with a contacting surface were meshed with second-order hexahedrons; the other partitions were meshed with second-order tetrahedrons to enable quick size transitions with high aspect ratio elements. Contact zones were meshed with second-order quadrilateral contact-target elements, allowing a penetration of 0.1 microns using augmented Lagrange formulation. The cross-sections illustrating the mesh of both models are shown in Fig. 3. The conventional bearing model has 321.813 DoF, whereas the wire bearing model has 614.529 DoF.

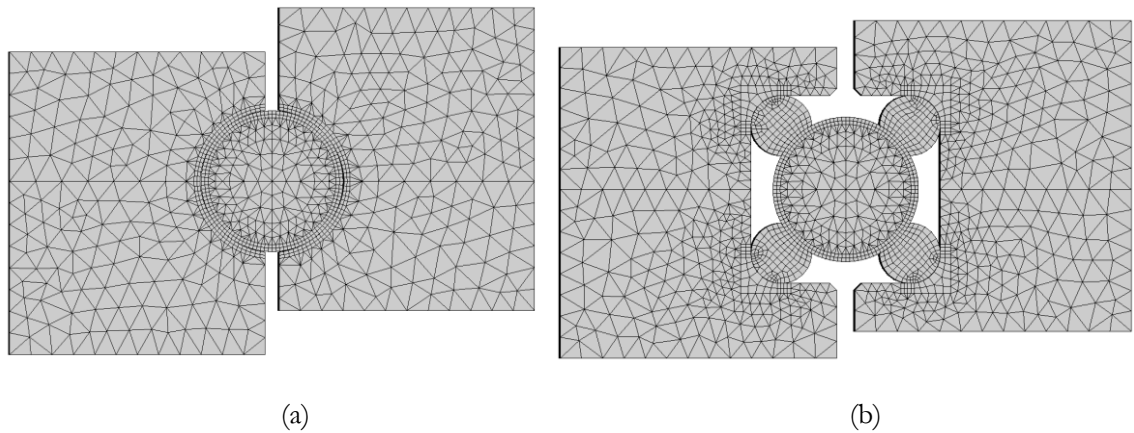

(a) (b)

**Fig. 3.** Cross-section of half sector models: (a) Conventional bearing (b) Wire bearing.

The external axial load pointed out in Table 1 was introduced by imposing an axial displacement to the upper surface of the outer ring in order to improve the convergence of the analyses. Besides, symmetry conditions were applied in the symmetry surfaces. Finally, for the clamped configuration, the lower face of the inner ring was fixed, thus restraining axial and radial movement; for the unclamped configuration, frictionless contact condition was imposed to the lower face of the inner ring, allowing for free radial movement.

### 2.2.2. Submodelling technique

Even though the global behaviour of the bearing can be accurately simulated by the half sector models, finer mesh is necessary in the contact surfaces in order to obtain better local contact results. In this sense, submodelling technique is a highly efficient tool [21]. In this case, the original models were the half sector models in Fig. 3, and the submodels were defined as the contact regions illustrated in Fig. 4 for both the conventional and the wire bearing. Thus, a first analysis is performed applying the axial displacement to the half sector model; then, the displacement results in the contact region boundaries are transferred to the submodel, analysing this partial geometry with a finer mesh and therefore with more accurate results. Due to the smaller dimensions of the submodels, finer contact meshes can be defined (2.416.227 DoF in the conventional bearing submodel, and 1.167.117 DoF in the wire bearing submodel) and consequently more accurate contact results can be obtained.

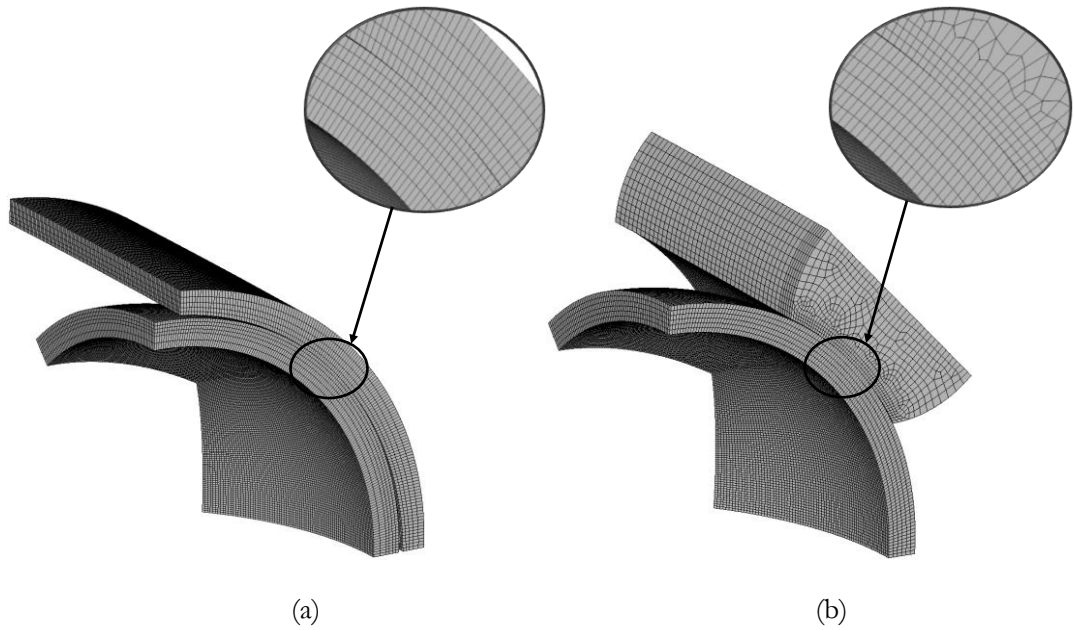

**Fig. 4.** Submodels: (a) Conventional bearing (b) Wire bearing.

## 3. Results and discussion

In this section the FE results are presented and discussed to draw the main conclusions of the work.

### 3.1. Wire twisting

The most remarkable phenomenon that takes place during the loading process is the twist of the wire. When the axial displacement is applied in conventional bearings, the ball climbs the raceway, increasing the ball-raceway contact angle. In wire bearings, the ball-wire contact force generated by the axial displacement is not aligned with the centre of the wire cross-section, and consequently a twisting moment is induced in the wire. Depending on the friction coefficient of the contacting surfaces, this moment promotes the wire twist rather than ball climbing, as in a conventional bearing. Fig. 5(a) shows the load case, Fig. 5(b) shows a detailed view of the undeformed mesh in the contact zone (note the coincident nodes along both wire-ring circumferential contact lines), and Fig. 5(c) illustrates wire twist as consequence of the applied load. The wire twisting has a huge influence on the behaviour of the wire bearing, due to its effects on the stiffness of the bearing, contact ellipse truncation and contact forces.

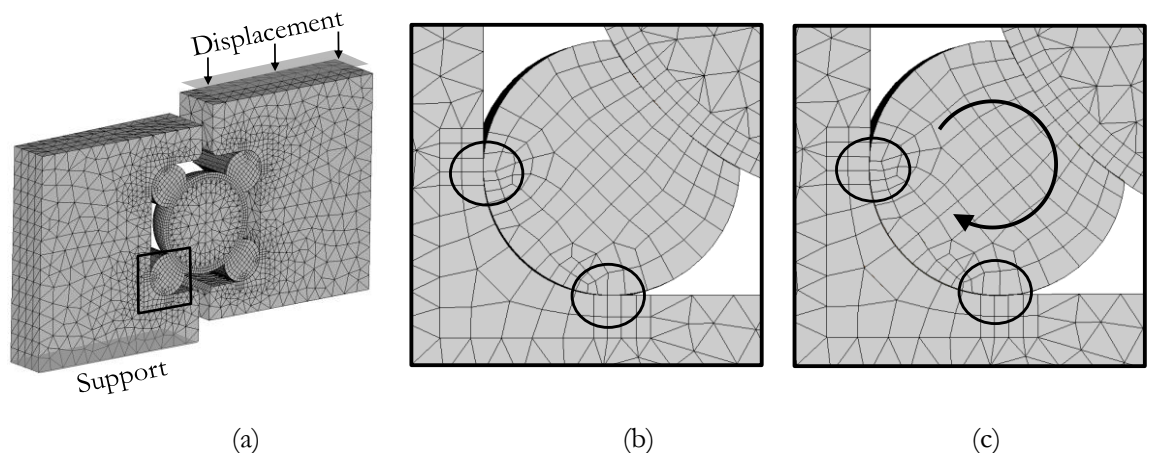

**Fig. 5.** Wire twist Case 3: (a) Load case (b) Undeformed model (c) Deformed model (scale x1.6).

## 3.2. Axial stiffness and static load capacity of the bearing

Axial stiffness curves for each case in Table 1 were obtained by means of FE analyses of the half sector models in Fig. 3, as the relationship between the displacement of the upper surface of the outer ring and the reaction forces in the lower surface of the inner ring. Moreover, the analytical model [7] used to fix the target axial force in Table 1, was also used to obtain the stiffness curves; this analytical model assumes rigid rings, so larger stiffness is expected. Fig. 6 shows the stiffness curves of the five cases summarized in Table 1, as well as the points in which the contact ellipse begins and completes truncation for unclamped (Fig. 6(a)) and clamped (Fig. 6(b)) situations. To this end, truncation was considered to begin when the contact ellipse reaches the raceway boundaries, and it was assumed to be complete when the maximum contact pressure was located at the boundary instead of in the centre of the contact ellipse. Next, the results of Fig. 6 are discussed.

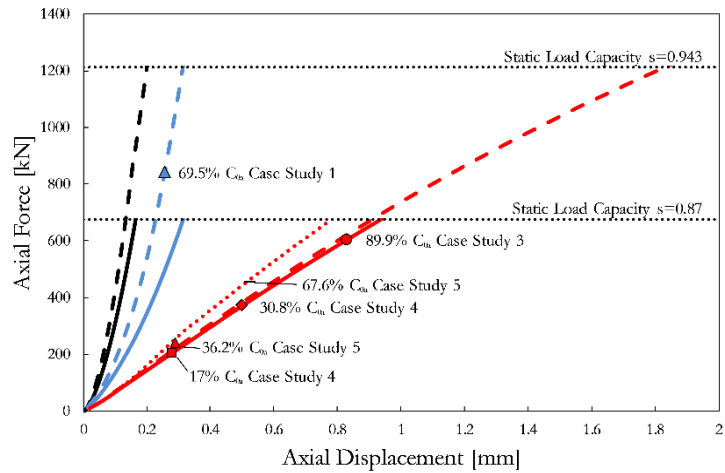

(a)

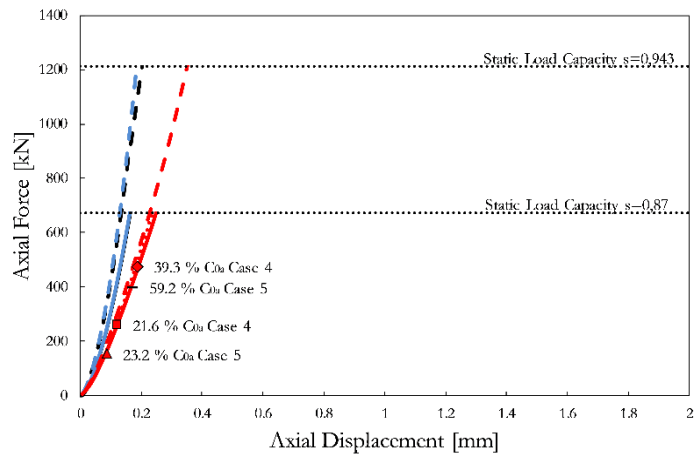

(b)

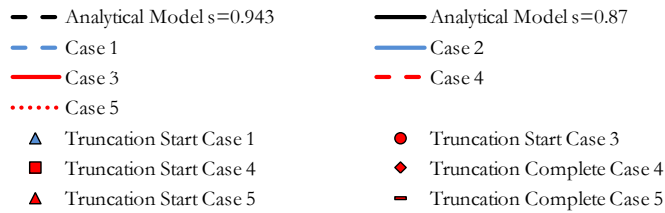

**Fig. 6.** Stiffness of the different bearings and truncation status: (a) Unclamped situation (b) Clamped situation.

**Axial stiffness behaviour**

It can be observed that the stiffness provided by the analytical model fits very well with the conventional bearing models with clamped configuration (Fig. 6(b)), and is slightly larger for the unclamped condition (cases 1 and 2 in Fig. 6(a)). This is because the clamped condition restricts the radial deformation of the rings, thus being closer to the rigid rings assumption of the analytical model.

Wire bearing rings are built with lighter and more compliant materials. This fact, together with the wire twisting, makes them more flexible than the rings of conventional bearings. The wire twisting effect on the stiffness can be clearly appreciated observing the wire bearings with the same conformity factor value (0.87) but different wire-ring friction coefficients (0.1 and 0.3), i.e. cases 3 and 5; a larger friction coefficient decreases the wire twisting and consequently increases the axial stiffness of the bearing.

Another phenomenon to take into account is the different stiffness behaviour of each bearing type under different boundary conditions. Conventional bearings have exponential stiffness behaviour due to the exponential nature of the ball-raceway contact deformation and the variation of the contact angle. Wire bearings provide almost linear stiffness behaviour for unclamped condition, mainly caused by the low stiffness of the rings and the slight variation of the contact angle due to the wire twisting, as it will be explained in the following section; for the clamped configuration, the flexibility of the rings does not play such an important role, and therefore the response is exponential.

**Static axial capacity and contact ellipse truncation**

According to Table 1, and as illustrated in Fig. 6, the static axial load capacity obtained from the analytical model [7] highly depends on the contact conformity: the most conformal bearings ($s$=0.943) have approximately twice the theoretical capacity of the less conformal ones ($s$=0.87). However, the analytical model does not consider the truncation of the contact ellipse, which can have a huge effect in the static capacity. From this point of view, Fig. 6(a) shows that for unclamped configuration, cases 1 and 3 have a similar behaviour: case 1, the conventional bearing with $s$=0.943, started suffering truncation at 69.5% of the theoretical static load capacity and did not reach the complete truncation, whereas case 3, the wire bearing with $s$=0.87 and $\mu$=0.1 in the raceway-ring contact, started suffering truncation at 89.9% of its theoretical static load capacity and neither reached the complete truncation.

As it has been mentioned in the previous section, as the wire-ring friction coefficient increases, so does the axial stiffness because the wire twist decreases, especially in the unclamped configuration; wire-ring friction also affects the contact ellipse truncation. The wire bearing of case 5 ($\mu$=0.3), starts truncation at 17%$C_{0a}$ and completes it at 67.6%$C_{0a}$, which is clearly worse than the response of case 3 ($\mu$=0.1). This fact demonstrates that in terms of static load capacity, the wire twisting phenomenon improves the performance since it prevents the truncation of the contact ellipse. In the clamped configuration, wire twisting is more restricted, so this phenomenon is not so critical.

Analysing the results of cases 2 and 4, both have clear disadvantages. On the one hand, the low conformity conventional bearing (case 2, with s=0.87) is less optimal than the high conformity one (case 1, with s=0.943) because it has half the static load capacity. On the other hand, the high conformity wire bearing (case 4, with s=0.943) started and completed the truncation of the contact ellipse at very low percentages of the static load capacity, both for clamped and unclamped conditions.

### 3.3. Contact forces and contact angles in the ball-raceway contact

The contact normal force ($Q$) is commonly used in analytical models to obtain the contact pressure and shear stress distribution [7,11,19]. The contact angle ($\alpha$) is defined as the angle of ($Q$) with the horizontal axis. In this work, the contact normal force was obtained by means of a postprocessing macro in Ansys®, based on the assumption that the normal contact force is the vector sum of the normal forces in each node; in order to validate this procedure, the angle between the point of the contact ellipse with the maximum pressure and the horizontal axis was measured, and it was found to be identical to the one obtained by means of forces. Thus, Fig. 7 illustrates the evolution of the ball-raceway contact angle with the normal force for each bearing under clamped and unclamped situations.

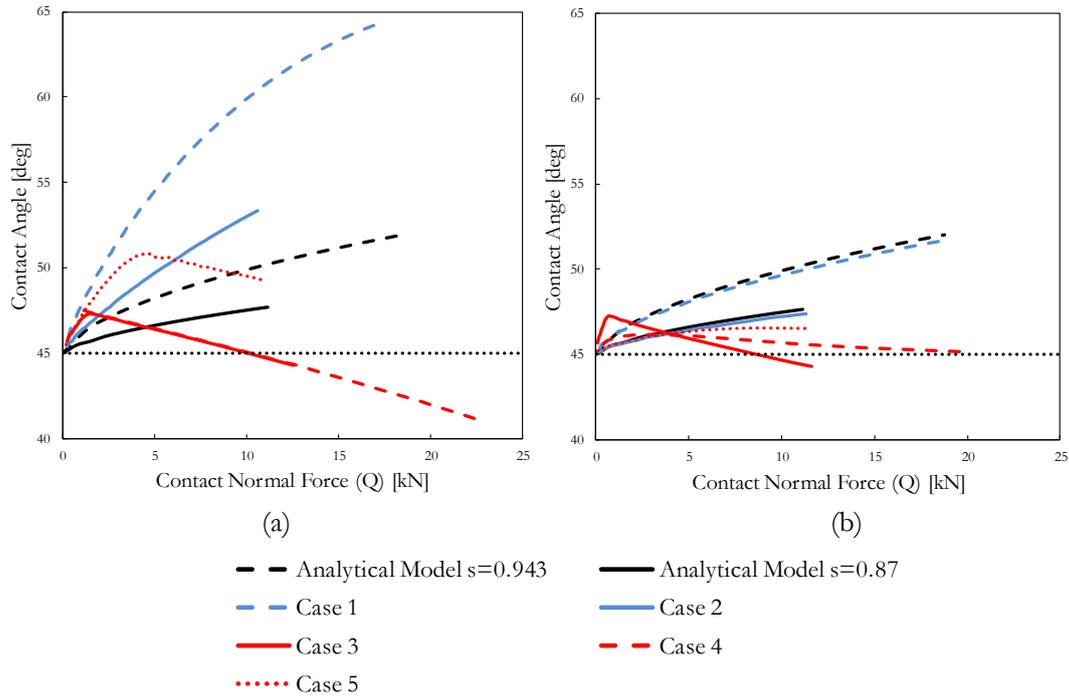

**Fig. 7.** Contact forces and angles a) Unclamped situation b) Clamped situation.

Fig. 7 shows that, in conventional bearings (cases 1 and 2), the contact angle increases with the axial load due to ball climbing, which can finally result in ball-raceway contact ellipse truncation. Thus, if the contact normal force and the contact angle are known, the dimensions of the contact ellipse can be calculated; if it reaches the limit of the raceway, truncation occurs. In wire bearings, wire twisting involves that the contact angle not always increases with the axial load, as illustrated in cases 3, 4 and 5 in Fig. 7: for low load values, the contact angle increases as in conventional bearings, because ball climbing is favored rather than wire twisting; however, from a given axial load on, wire twisting starts and consequently contact angle decreases. Due to this complex behavior, the study of the contact ellipse truncation in wire bearings is not as straightforward as in conventional bearings. As the schematic illustration in Fig. 8 shows, wire twisting involves contact angle decrease but not contact ellipse truncation, because the contact ellipse remains centered in the wire raceway. This statement is demonstrated by the plots in Fig. 9, which shows the contact pressure distribution along the major semi-axis of the ellipse for increasing load values: for conventional bearings (case 1 in Fig. 9(a), and case 2 in Fig. 9(b)), contact ellipse moves towards the raceway limits as the load and consequently the contact angle increases (see cases 1 and 2 in Fig. 7); on the contrary, for the wire bearing of case 3 with unclamped condition, Fig. 9(c) evinces that the contact ellipse remains centered, even though Fig. 7(a) shows that the contact angle clearly decreases according to Fig. 8; finally, if the wire-ring friction coefficient is increased (case 5), Figs. 7(a) and 9(d) show that a larger load is needed to start wire twisting, which initially favors ball climbing and therefore contact ellipse truncation. As a consequence, ball-raceway contact angle alone is not enough to study contact ellipse truncation in wire bearings, wire twisting must also be taken into account; this aspect is especially critical if simplified analytical models are to be developed. Fig. 10 shows the contact ellipses of the four plots in Fig. 9 for the 100% of the target axial load (static load capacity). Note that, according to Fig. 9, the maximum pressure is not exactly 4200 MPa for that load value, as it should be. The static load capacity in Table 1 was calculated using the analytical model, which considers rigid rings; as the rings are flexible in the FE model, the contact angles and normal forces are slightly different from those predicted by the analytical model, and so is the maximum contact pressure.

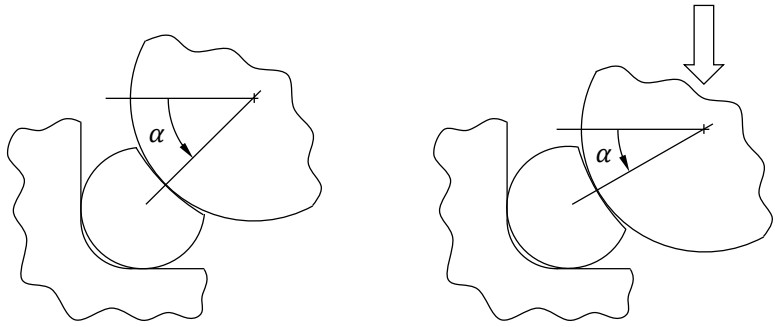

**Fig. 8.** Structural response of the wire bearing under axial load.

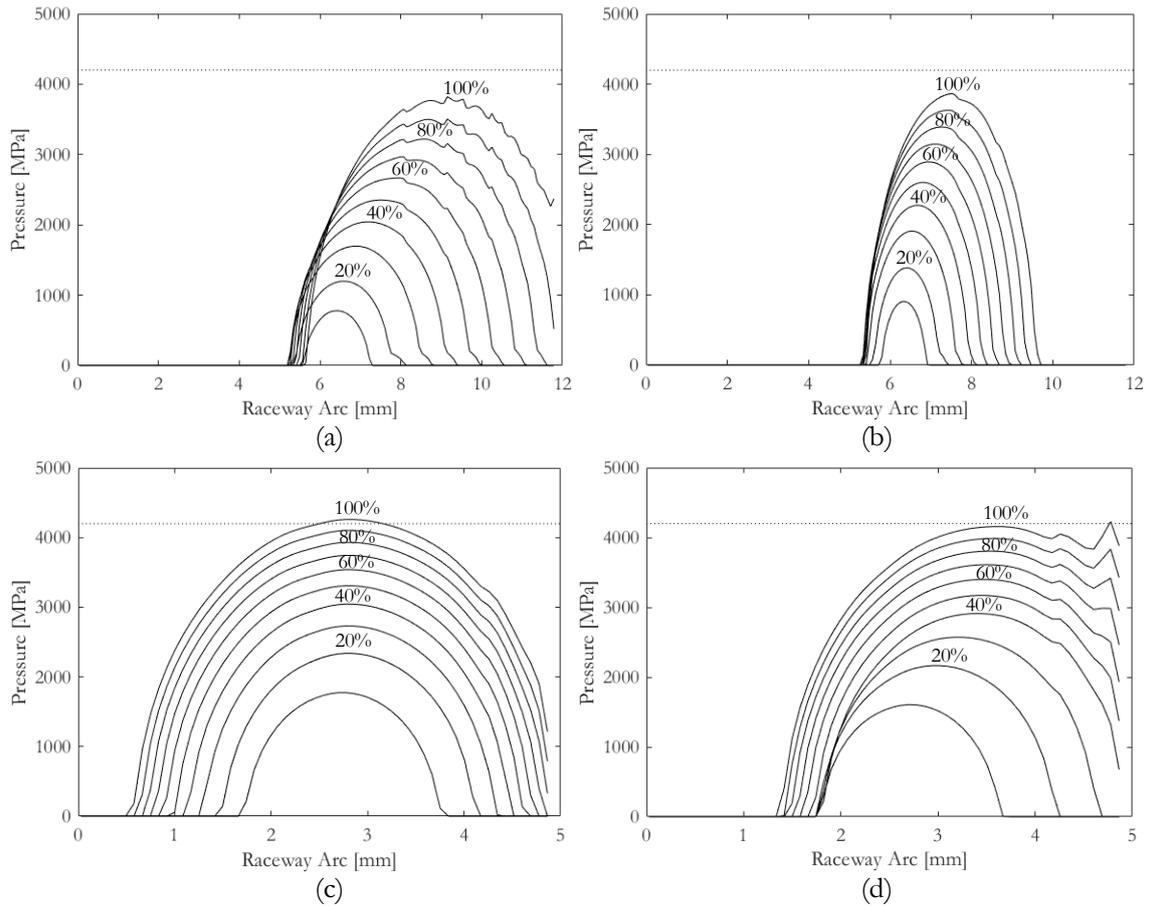

**Fig. 9.** Evolution of the pressure in the contact of the symmetry plane during the loading process (unclamped conditions): (a) Case 1 (b) Case 2 (c) Case 3 (d) Case 5.

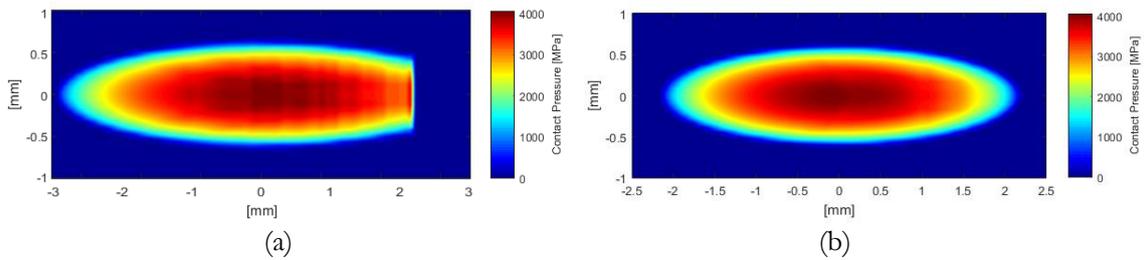

(a)                                                      (b)

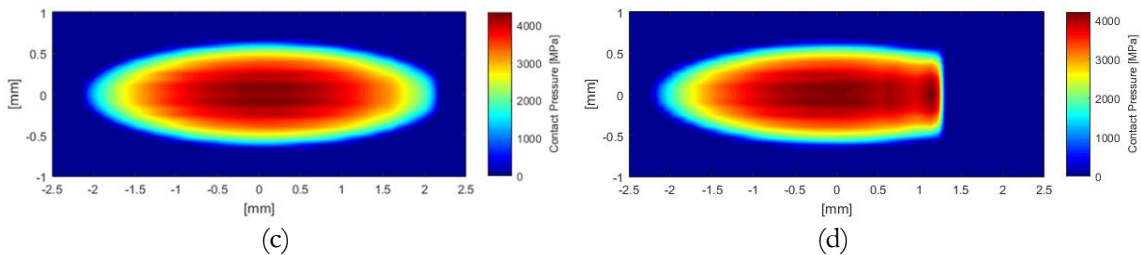

(c)                  (d)

**Fig. 10.** Pressure contact ellipse 100% of the target axial load (unclamped conditions): (a) Case 1 (b) Case 2 (c) Case 3 (d) Case 5.

## 4. Conclusions

Little journal publications have been released about the mechanical behaviour of wire bearings. This work compares four-point contact wire bearings with conventional bearings of the same type using Finite Element models. For such purpose, static axial load response is analysed under different design and operational conditions, such as the contact conformity, friction coefficient, and the stiffness of the supports. The most remarkable feature in wire bearings, apart from the flexibility of the rings (made of aluminium, instead of steel as in conventional bearings) is the wire twisting phenomenon, i.e. the rotation of the wire cross-section under external load. Wire twisting increases as the wire-ring friction coefficient or the stiffness of the supports decreases. For these reasons, wire bearings are more flexible than conventional bearings. Regarding local ball-raceway contact results, even though wire raceways are shorter, wire twisting inhibits ball climbing; instead of that, wire section rotates keeping the contact ellipse centred in the raceway; as a result, contact ellipse truncation is not a major problem; in this sense, within the limitations of this study, wire bearings with a low conformity value behave similarly to conventional bearings in this aspect. The wide variety of cases studied in this work have demonstrated that the response of wire bearings can easily be adjusted to achieve the desired working conditions by controlling variables such as the wire ring friction coefficient by means of lubrication, the ball-raceway conformity value, or the stiffness of the supporting structures.

## Acknowledgements

The authors want to acknowledge the financial support of the Spanish Ministry of Economy and Competitiveness through grant number DPI2017-85487-R (AEI/FEDER, UE); the Basque Government through project number IT947-16; and the Basque Bearing Manufacturer Iraundi Bearings S.A. for the help in the geometrical definition of the wires and many background items of the research.